\documentclass[preprint]{aastex}
\usepackage{graphicx}
\usepackage{natbib}
\usepackage{amssymb}
\usepackage{subfigure}
\bibpunct{(}{)}{;}{a}{}{,} 

\shortauthors{Li \& Smith}
\shorttitle{Cluster formation in the south-east boundary of the RMC}

\begin{document}

\title{Multi-seeded multi-mode formation of embedded clusters in the RMC: 
Structured star formation toward the south-east boundary}
\author{J. Z. Li$^{1,2}$ \& M. D. Smith$^{2}$}
\affil{$^{1}$National Astronomical Observatories, Chinese Academy of 
           Sciences, 20A Datun Road, Chaoyang District, Beijing 100012, China; ljz@bao.ac.cn \\
      $^{2}$Armagh Observatory, College Hill, Armagh BT61 9DG, N. Ireland, UK; mds@arm.ac.uk}

\begin{abstract}

The Rosette Molecular Complex contains embedded clusters with diverse
properties and origins. We have previously explored the shell mode of formation 
in the north (Regions A \& B) and the massive concentrations in the ridge (Region C).
Here, we explore star formation towards the south of the complex, Region D,
based on data from the spatially complete 2 Micron All Sky Survey. We find that
stars are forming prolifically throughout this region in a highly structured mode 
with both clusters and loose aggregates detected. The most prominent cluster (Region D1) 
lies in the north-center. This cluster is over 20~pc to the south of the Monoceros 
ridge, the interface of the emerging young OB cluster NGC~2244 with its ambient molecular 
clouds. In addition, there are several branches  stemming from AFGL~961 in Region~C
and extending to the south-east boundary of the cloud.
We invoke a tree model to interpret this pattern, corresponding to probable tracks of 
abrupt turbulent excitation and subsequent decay. Alternatively, we discuss gravoturbulent
collapse scenarios based on numerical simulations. Relative stellar ages and gas flow
directions will differentiate between these mechanisms. 

\end{abstract}


\section{Introduction}

The Rosette Molecular Cloud (RMC)
is known as one of the most massive giant molecular 
complexes (GMCs) in the Galaxy \citep{1980ApJ...241..676B}. At a distance of $\sim$~1.5~kpc, 
it is among the most prominent sites for the exploration of active star formation 
\citep{1990A&A...230..181C,1993A&A...273L..41B}.  Indeed, the presence 
of several clusters of young stars associated with massive clumps were recognised 
on near-infrared K-band images \citep{1997ApJ...477..176P}. \citet{2005LSa} 
were able to probe the clusters in more detail by using 
the three-band spatially complete 2 Micron All Sky Survey (2MASS). 
They showed that deeply-embedded clusters, sub-clusters
and loose aggregates are widely distributed throughout the RMC. 

Cluster formation in the RMC is taking place in a multi-seeded manner i.e. clusters are 
forming in parallel yet in independent locations \citep{2005LSa}. 
There is also strong evidence for multi-mode formation, with the mode related to
the cloud location. Four regions were identified by \citet{2005LSa} as also indicated here in
the upper panel of Fig.~\ref{structure}.
 
Firstly, Regions A \& B were identified with the interaction layer of the  Rosette Nebula
with swept-up and fragmented arcs of molecular gas \citep{2005LSb}. The interface contains several
loose aggregates spread along the compressed cloud layers. In Region~C, a burst 
of formation of compact sub-clusters and high-mass stellar groups, signifying new generations 
of OB cluster formation, is taking place \citep{2005LSc}.
Region~C is associated with the densest ridge of the RMC.

Star formation in the RMC that has previously taken place 
\citep{1964ARA&A...2..213B,1979Ap&SS..66..191S,1976ApJ...210...65T}
or is now proceeding \citep{2005LSa} could well be consistent with the scenario of sequential 
formation of massive clusters in GMCs \citep{1977ApJ...214..725E,1987IAUS..115....1L}.
If the choice is between spontaneous and triggered collapse, then the latter has considerable
support. However, the specific modes and 
mechanisms to be held responsible in the contrasting areas remain to be
determined.  In this paper, we will focus on the  embedded population 
of young stars in Region D, toward the south-east boundary of the RMC.
At a distance exceeding 20\,pc from the interaction layer, the influence from 
the emerging young open cluster NGC~2244 and its associated H{\small II} region is 
very limited \citep{1990A&A...230..181C,1998A&A...335.1049S}.
On the other hand, being in the vicinity of the massive young stars in Region~C,
regions of rapid dispersal and strong compression might be evident.
Therefore, exploration of 
Region~D may offer a critical test of the conventional understanding of sequential 
cluster formation \citep{1987IAUS..115....1L}. 

\section{Data Acquisition \& Analysis}
A detailed review of the methodology was presented in the first of 
this series \citep{2005LSb}.
We employ color constraints to substantiate the embedded nature of the clusters.
Control fields were utilised to help determine the proportions of embedded
and background stars. 

Both archived data from the 2MASS Point Source Catalogue (PSC) and the IRAS Sky 
Survey Atlases (ISSA) were retrieved via IRSA (http://irsa.ipac.caltech.edu/). 
For a detailed introduction to the 2MASS mission and the 2MASS
PSC, please consult the 2MASS Explanatory Supplement.  The 2MASS PSC 
contains 470 million source extractions with a broad range of photometric 
quality. Restrictions to the 2MASS photometric data are necessary 
to meet the requirements of different studies. Here, we extract only those sources
with certain detections in all three near infrared bands: J, H and Ks.
We then further restrict the Ks band signal to noise ratio to be above 15, to
constrain stars in the control fields to the unreddened main-sequence
and post-main-sequence tracks on the (J-H) to (H-Ks) diagram (see 
Fig.~4a of \citet{2005LSb}).
 
\section{Spatial distribution of embedded clusters and loose aggregates}

The spatial distribution of the Region~D sample sources are plotted as a function of 
H-Ks color in Fig.~\ref{spatial}. The color distinction helps to eliminate a possible random
distribution of foreground field stars without excluding potential cloud members which may also be
widely distributed.
Reddened sources with a H-Ks color above 0.5 and 0.7 are overplotted onto the distribution 
of optical depth at 100 $\mu$m (upper panel) and cold dust temperature (lower panel), 
respectively, as derived from the ISSA data \citep{1996ChA&A..20..445L}. 
Clustering and fine structure in the distribution 
begin to appear as the minimum  H-Ks reddening is increased.

A dense cluster is readily seen to the north of the center of this region, Region D1. 
This sub-cluster possesses a symmetric and compact structure, as does the associated 
molecular clump \citep{1980ApJ...241..676B,1995ApJ...451..252W}.
We could expect D1 to be the result of largely spontaneous or non-triggered cluster formation, 
rather than having been primarily induced by the expanding Rosette Nebula and its embedded 
OB cluster NGC~2244. The spatial separation of more than 20~pc between Region D1 
and the Monoceros ridge provides support for this argument, consistent with the claims
made by \citet{1980ApJ...241..676B}.
In addition, ultraviolet ionization of the inter-clump gas is probably insufficient 
to induce implosion of the clumps at and beyond this distance 
\citep{1980ApJ...241..676B,1990A&A...230..181C,1998A&A...335.1049S} despite
the large clump to inter-clump density contrast detected in the RMC 
\citep{1980ApJ...241..676B,1995ApJ...451..252W}. This issue will be further
elaborated in \S~${7}$.

We cannot state unequivocally which of the reddened stars are embedded and 
which are background through just the stellar distribution. However, several lanes of
reddened sources and widely distributed aggregates 
are discernible in Region D. In Fig.~\ref{spatial}, these regions are outlined
by either circles or rectangles according to their appearance.  
Some of the stellar concentrations are known to be associated with IRAS sources
\citep{1990A&A...230..181C} and young stellar groups \citep{1997ApJ...477..176P}. 
This provides first evidence that the majority of the reddened sources associated with
the regulated structures are embedded in the cloud, rather than
resulting from reddening of background stars.
Nevertheless, in areas of clustering other than D1, no major gas clumps
as prominent as those located in Region C exist \citep{1995ApJ...451..252W},
as also indicated by the optical depth distribution (upper panel).
Some aggregates can be traced to regions with comparatively low opacity, beyond
the filaments in the extension of the RMC detected in lines of CO \& C$^{13}$O
\citep{1995ApJ...451..252W}.
This further reduces the attribution of the clustered appearance of the reddened 
sources in Region D to the effects of cloud extinction and reddening on randomly 
distributed background stars.
On the other hand, certain stellar groups are associated with relatively warm
dust temperatures of 25\,-\,41~K (lower panel), which may be a 
factor related to their gestation or to the heating by massive embedded stars.

To measure the significance of the clustering of the candidate young stars 
in Region D and particularly those associated with Region D1, we have computed the standard 
two-point correlation function. We first determine the number of stellar
pairs, $H_d({\theta})$, separated by an angle $\theta$, thrown into
logarithmic bins between $\log ({\theta}))$ and  $\log ({\theta})) + d \log ({\theta}))$.
This is compared to the predicted distribution for a random sample of
stars spread over the same area, $H_r({\theta})$. The two-point
correlation function is then defined as
\begin{equation}
\Phi({\theta}) = \frac{H_d({\theta})}{H_r({\theta})}  -   1.
\end{equation}
Hence, a random distribution yields $\Phi = 0$. Such a
random distribution is found in Fig.~\ref{correlation} only
when we take all the stars with H-Ks~${>}$~0.2~mag in the entire Region~D
(dashed line).

As demonstrated in Fig.~\ref{correlation}, the higher
the (H-Ks) color constraint placed on the sample of sources, the stronger
the resultant spatial correlation and, hence, the clustering of sources. This holds
for both the entire region and the sub-cluster (Fig.~\ref{correlation1}). In fact,
the values of $\Phi$ uncovered here are significantly larger than found in other
clusters in the Rosette Complex (maximum values of $\sim$ 6 (A), 0.5 (B) and 1.5 (C)
were found).

Two important results emerge on re-plotting the number $\Phi$
logarithmically, as shown in Fig.~\ref{corrlog}. Firstly, the
correlation functions are consistent with power-laws.
This implies that {\em there is no particular scale length which we could identify
with either a Jeans length or other fragmentation scale}.

Secondly, the correlation function is steep. The $\Phi$--$\theta$ power-law index
is $\gamma \sim -0.84$ for the samples with H-Ks~${>}$~0.7~mag in Region~D
and   $\gamma \sim -0.95$ in Region~D1, as indicated by the dot-dashed line segments.
The function is steep in comparison both to other regions of the RMC \citep{2005LSb,2005LSc}  and to
other clusters \citep[e.g.][]{1999ApJ...510..258S}. Hence, clustering is particularly strong here
suggesting that these stars are {\em either tightly bound or extremely young}.
We can ask if the strong correlation on the fine scales may be partly due to the distribution of
sources into the branches discussed below. However, it is clear that the branches 
are quite wide (of order of 0.2$^\circ$) and we find no particular 
length scale that could be associated with this width.

The distribution of projected separation distances between nearest neighbours contains
distinct information.
While the two-point correlation function describes the nature of the hierarchical
clustering, the nearest neighbour distribution will emphasise the
presence of any scale of fragmentation or a scale associated with gravitational
confinement. The left panel of Fig.~\ref{neighbours} demonstrates an almost perfect agreement
with a random distribution for the entire sample with H--K exceeding 0.2.
In contrast, for the deeply embedded stars (the right panel), there is a significant  excess
of neighbouring stars with a separation of $\sim$~30\arcsec and a second possible
excess at $\sim$~55\arcsec. Given the broad distribution expected for a random sample,
this is more plausibly interpreted as a deficit of separations of  $\sim$~40--45\arcsec,
corresponding to a scale of 0.3~parsec. Given the size of the error bars, however, we just note 
here that there is an indication of a scale length which may correspond to that of clump 
fragmentation into massive stellar binaries.
  
\section{Color-Color Diagrams}

To explore the nature of the candidate young stars in Region D and particularly Region D1,
all 2MASS sources that match the sample selection criteria introduced
in \S~${2}$ are put onto (J--H)---(H--Ks) diagrams (Fig.~\ref{colorcolor}). 
Color-color diagrams for two off-cloud control fields were constructed as
a comparison \citep{2005LSb} which indicate negligible foreground
and background extinction generally toward the RMC field. 

Considerable fractions of the reddened sources
are found to be located to the right of the reddening
band of normal field stars in the Region D diagram. The
fractions are nearly one third and a half for the sources with H-Ks above
0.5 and 0.7 mag, respectively. The infrared excesses indicate their youth, 
and provide support that the very young stars in Region D 
are associated with the molecular cloud.

Nevertheless, contamination by background field stars in this region is
high due to the comparatively low extinction at the south-east boundary
of the RMC \citep{1980ApJ...241..676B,1995ApJ...451..252W}
and the large sky area involved. Indeed, a clear
concentration of reddened sources are located to the left edge of the reddening 
band and these are probably not cloud members. 
Due to their position in the reddening band, the majority of these sources 
could well be background giants. The visual extinction in this region is found
to range from $\sim$ 0.5 to 24.5~mag and indicates a mean of $\sim$~4\,mag.  

The most extreme source in the Region~D diagram is identified as a
heavily reddened source invisible in the optical. Its position in the 
diagram suggests a pre-main sequence origin of the object.

Inspection of the color-color diagram for Region D1 reveals
no clear concentration of reddened sources in the reddening band. About 20\% of the reddened
sources are located to the right of the reddening band for normal stars
and are probably intermediate and low mass young stars with intrinsic
excess emission.  The sub-cluster shows a maximum 
visual extinction of roughly 20\,mag. However, we find no evidence for protostars
with high infrared excesses in this sub-cluster, indicating that very massive stars
are not currently forming in this specific region. 

\section{Color-Magnitude Diagrams}

Color-Magnitude Diagrams (CMD) are presented in Fig.~\ref{colormag}.
The near-vertical solid line represents the loci of stars on the unreddened main-sequence.
A clear separation between the foreground main sequence stars and the
candidate cluster members is apparent. 

About one sixth of the candidate cluster members 
are located above the reddening vector drawn for a A0 dwarf. This
indicates that a considerable number of sources might be massive protostars
still deeply immersed in their parental cloud cores or envelopes. The location
of the most extreme source among these is indicated by a diamond. It can be a pre-main
sequence star as massive as $\sim$~50 M$\odot$ and is the most massive candidate young
stellar object in this region. Region D is thus associated with a structured distribution of 
loose aggregates containing massive young stars.

The compact sub-cluster residing in Region D1, as indicated by the
CMD, is probably a medium mass embedded cluster. The most massive
candidate young stars have ultimate masses of below or around 20~M$\odot$.
Nearly one third of the candidate cluster members in this region
are located above the reddening vector of a A0 dwarf and are good
candidates for intermediate mass young stars.


Some sources are easily detected in the H and Ks bands but are overlooked in the J band,
mainly due to the sensitivity of the J-band detection to extinction along the line of sight
\citep{2004A&A...421..623K}. Other possible reasons for exclusion include (i) being 
shrouded by apparently bright sources associated with or nearby in projection and 
(ii) severely obscured from identification by diffuse emission. 
Therefore, the strict sample selection method employed, along with the flux and resolution 
limited 2MASS survey, directly result in the omission of many sources well measured
in only the H and Ks bands. These sources were also accumulated and are presented on the CMDs 
of Regions D and D1 (plus signs).  This shows evidence that 
these criteria missed sources are predominantly candidates for 
embedded young stars suffering high extinction. High-resolution,
high sensitivity surveys are therefore required to perform a complete census of
the embedded population. 

\section{Ks Luminosity Function}

The luminosity function of embedded clusters is usually employed as a
tool to address questions related to the initial mass function.
The Ks luminosity functions (KLF) for Region D and Sub-region D1 are presented 
in Fig.~\ref{lumin}.  Both KLFs correspond
well to a power law distribution in the range of 11.0~${<}$~Ks~${<}$~14.5 
mag. They show a consistent slope of $\sim$~0.4 and, consequently, suggest 
a young age of $<$~1\,Myr when compared with 
embedded clusters of known ages \citep{1995AJ....109.1682L}. The embedded cluster
in Region D thus has an indicated age comparable to that of the Region C cluster.

A comparison of the KLF of each well-confined region of star 
formation in the RMC provides a consistent view of sequential star and cluster formation. 
The young open cluster NGC\,2244 with an age of around 2~Myrs is acknowledged to represent 
the oldest episode of recent cluster formation in the RMC \citep{2005L}. 
Clusters in the swept-up shells of the expanding H{\small II} region are subsequently 
formed, possessing ages of $>$ 1 Myr and are still closely associated with their birth sites
\citep{2005LSb}. Embedded clusters hatched in both Regions C \& D have KLF-indicated  
ages of the order of $<$~1\,Myr and so are among the most recently formed. However, 
although there is an age sequence from north to south,
it does not necessarily mean that the Region C \& D clusters are exclusively triggered 
by NGC\,2244, as we now discuss. 

\section{Schematic models}

An entire view of the reddened stars in the RMC is provided in Fig.~\ref{structure}a.
The distribution of the sub-clusters in Regions C \& D surprisingly resembles
an `upside-down tree' that is rooted at the center of Region~C. There are four
prominent branches traced with solid lines.
Remarkably, the IRAS 60\,$\mu$m map of this region (Fig.~\ref{structure}b)
also displays extensions in flux with a similar pattern in accordance with the
regulated distribution of the reddened sources. We can exclude its 
origin as due to monotone tracking problems of the IRAS survey since the flux pattern of dust
emission is elongated and oriented in different directions.

There are at least three potential causes for the structured star formation in Region D.
Firstly, in the {\bf internal sequential} scenario, the emerging young OB cluster 
in the north, NGC\,2244, triggers new generations of massive star formation in the 
densest ridge of the RMC \citep{2005LSc}.
These stars then trigger further generations, finally reaching the south-eastern
tip of the cloud.  Note that the trigger mechanism is clearly compression associated with the
H{\small II} region for clusters A \& B embedded in the interface but the impact of 
shocks further into the cloud should be severely impeded by the swept-up shells \citep{2005LSb}.
Deep into the molecular cloud, radiative implosion has been discussed as a possible mechanism by
\citet{1998A&A...335.1049S}, who find that the incident far-ultraviolet flux
is low in the RMC as compared to other star formation regions, though its partial
contribution to the active star formation in the south-east of the RMC cannot be
excluded. 

On the other hand, \citet{1990A&A...230..181C} present evidence that radio continuum emission
extends almost a degree to the south from the main cluster
NGC\,2244 in the direction from the Monoceros ridge toward AFGL\,961. This indicates that
the cloud ridge could be compressed by the H{<small II} region from behind the cloud.
However, the radio emission overlapping with AFGL\,961 is extremely weak
and absent further to the south. It is difficult to argue that ionization 
associated with the H{<small II} region could significantly raise the pressure
and be the primary trigger of the collapse of molecular clumps
associated with the compact clusters in Region D. 

Secondly, {\bf external triggering} should be considered.
The compression, fragmentation and collapse may be associated with
an old supernova or with the Perseus Arm \citep{1984SvAL...10..309G,1991RMxAA..22...99P}. 
The branches then represent the compressed
filaments which follow the decay of macroturbulence. On the other hand, Region D sits at the 
base of the 'relic' tail of the RMC that may have resulted from tidal force dissipation 
during its interaction with the Galactic plane.
This can be the reason why clustered star formation is taking place 
in this specific region along the Perseus Arm, where the general gas density is comparatively
low. This could also be related to how the tree-like structure and, consequently, 
the Region D sub-clusters and loose aggregates were formed.  

Once externally triggered, a giant cloud is expected to undergo {\em gravoturbulent collapse}
\citep{2001ApJ...549..386K}.
This implies that the cloud material collapses into sheets. Flow within the sheets leads to the 
formation of filaments. Mass accumulates where the filaments intersect, enhanced by the self-gravity.
Very high mass stars may then form in the high mass clump at the intersection. 
Other massive stars form within the dense filaments. As the intersection location
moves, new adjacent clusters containing high mass stars form. This global collapse should produce
the youngest massive stars in the central agglomeration, consistent with the observed state of the RMC.

Finally, we consider a {\bf tree model}, in which star formation proceeds
from the inside, out along the branches, as an alternative mechanism. The formation of 
the  structure is interpreted as directly triggered by the proto-O triple system
at the south-center of Region~C \citep{2005LSc}, the location of which seems to match well 
the root of the tree. Note that an additional branch of the tree structure seems
to exist extending from the massive multiple system associated with AFGL~961 to the 
eastern edge. Both the compact sub-clusters in Region~C and the loose aggregates in Region~D can 
be triggered by the violent activity related to the formation of the protostellar O stars.
Thus, the tree structure is here attributed to strong activity associated with
the high-mass star formation in Region C. The induced large-scale turbulence compresses
the edges of the cloud and then decays along radial directions, leading to systematic star 
formation along sheet-like structures. 

Nevertheless, several less conspicuous far infrared emission cores also exist and 
are extensively scattered within Region D. They have locations commensurate with 
the distribution of molecular clumps or filaments \citep{1995ApJ...451..252W}. 
Each is more or less spatially associated with a loose aggregate of reddened young 
stars, providing support to our remarks that recent clustered star formation 
is occurring in this region. However, the structured distribution of candidate 
young stars does not end where the clumps or filaments do. This suggests
that the tracks of star formation correspond to locations where the turbulence
has decayed and the gas has been efficiently transformed into stars.

\section{Summary}

Reddened stars are distributed within  clusters, loose aggregates and filaments
towards the south-east boundary of the RMC. The clustering is relatively strong in
Region D, as measured by the two-point correlation function. This suggest that the
stars have not had time to disperse, possibly commensurate with a very
young age, as testified by the KLFs. Supposing a velocity dispersion
of 2~km~s$^{-1}$, the stars would disperse within a region of size 4~pc after
1~Myr. At a distance of 1.5~kpc, this corresponds to 0.16$^\circ$. This is
roughly consistent with the widths of the branches along which the reddened stars
are aligned. After 2~Myr, however, the branches would be hard to discern.

Schematic models are presented which help elucidate the nature of clustered star formation 
throughout the south-east quadrant of the RMC. A scenario of extensive 
clustered star formation induced by the violent activity of the newly 
hatched proto-O stellar group associated with AFGL~961, from where the
branches appear to stem, is suggested. This tree model, in which the central roots
form before the branches, contrasts with the
gravoturbulent model, in which material flows inwards along sheets and filaments,
accumulating in the massive central ridge.

How can we differentiate between these two contrasting models?
Firstly, we can explore the gas flows within the RMC to gain evidence for inflow
or outflow. At present, it is known that turbulent motions dominate the dynamics,
with the addition of a  moderate large-scale velocity gradient 
\citep{1995ApJ...451..252W}. Secondly, we could derive more accurate ages for 
the young stars. If, as supported by the presence of two powerful molecular outflows
\citep{1998A&A...335.1049S}, the massive clusters along the ridge are indeed the youngest,
then the gravoturbulent model would be difficult to challenge.


{\flushleft \bf Acknowledgments~}

Beside the 2MASS Archive, this work also made use of IRAS PSC \& ISSA data.
DSS Survey data (STSCI, funded by NSF) were also employed. This project is
sponsored by SRF for ROCS, SEM and is partially supported by the Department
of Culture, Arts and Leisure (Northern Ireland) and PPARC (U.K.).



\clearpage
\figcaption[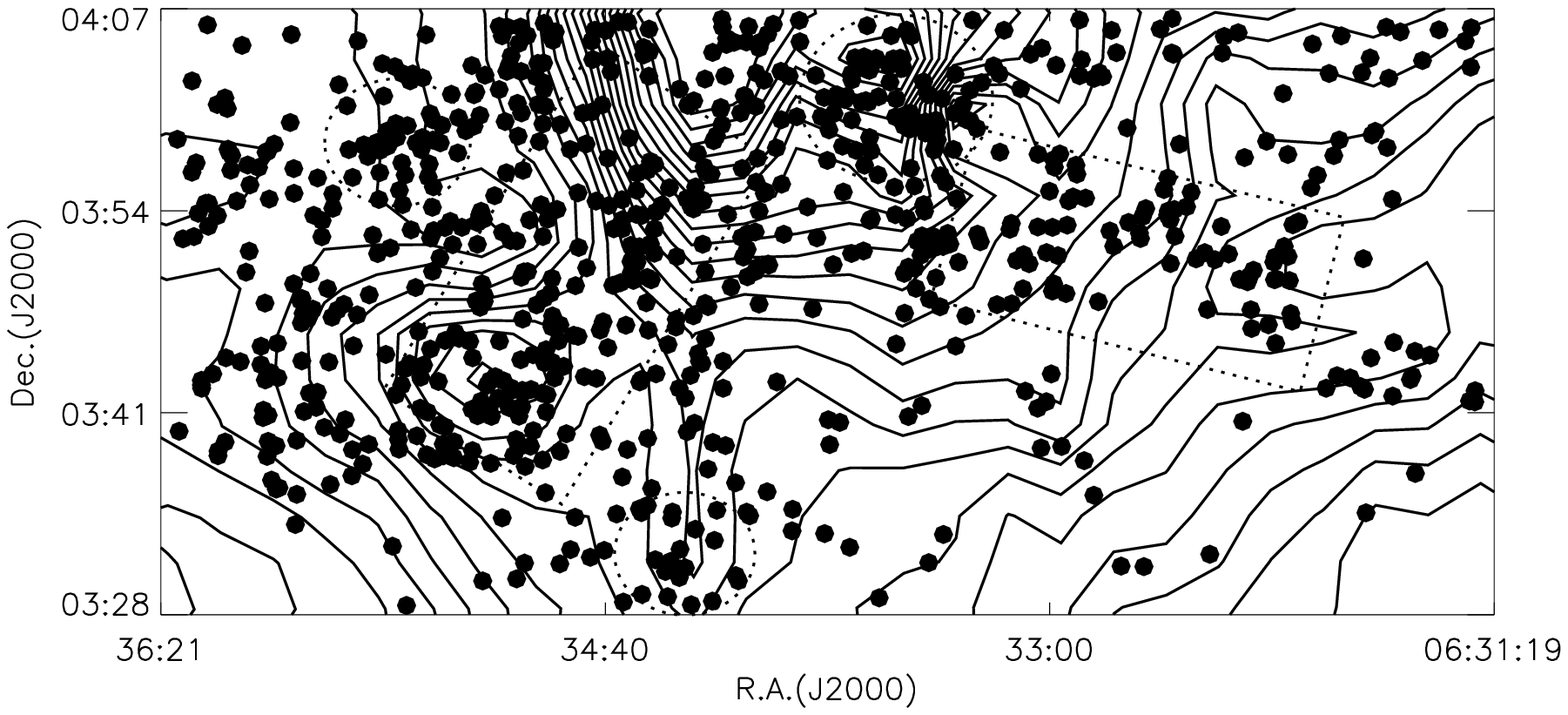,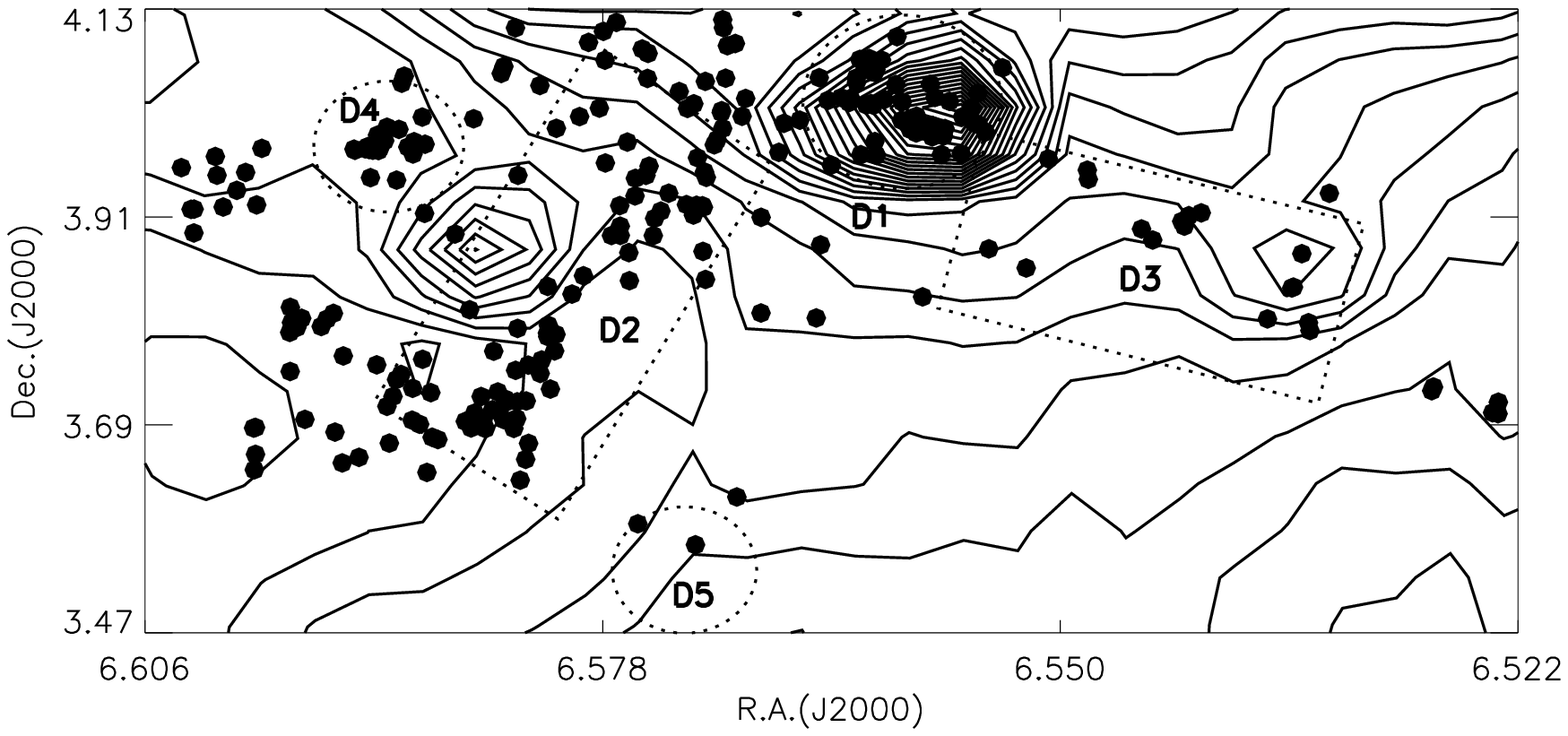]{The spatial distribution of the Region D sub-clusters and aggregates with 
(H-Ks)$>$~0.5 (upper panel) and $>$~0.7 (lower panel), respectively. Note that the
source distribution of reddened sources is overplotted onto the distribution of optical depth
at 100\,$\mu$m in the upper panel and onto that of the color-temperature in the lower 
panel.  The distribution of optical depth and cold dust temperature was calculated 
from IRAS 60\,$\mu$m and 100\,$\mu$m images \citep{1996ChA&A..20..445L}. \label{spatial}}

\figcaption[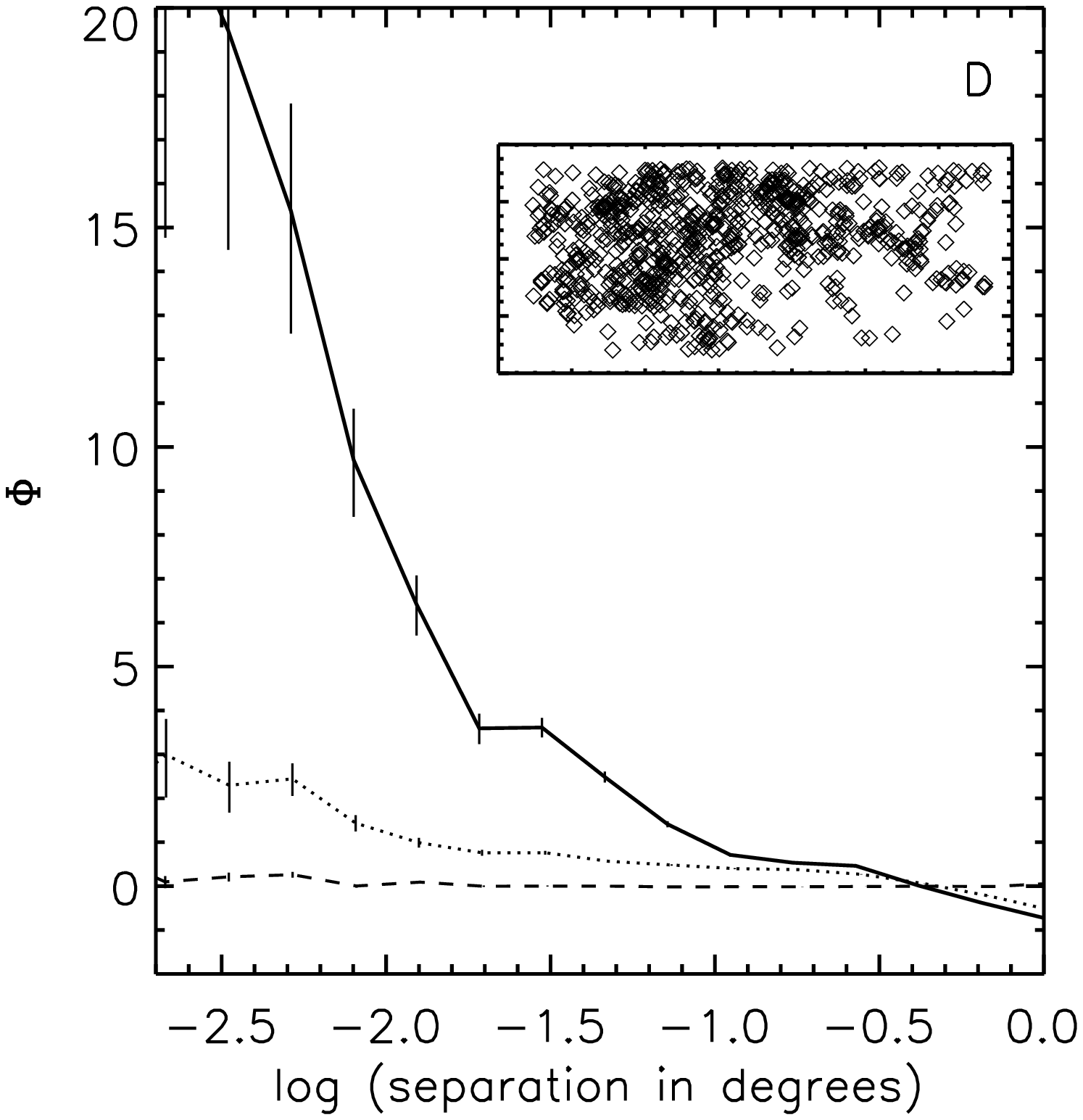]{The two-point correlation function for objects in 
Region~D of the RMC.
The three lines correspond to the data with  H-Ks~${>}$~0.2~mag (dashed line, 2660
stars), H-Ks~${>}$~0.5~mag (dotted line, 792) and  H-Ks~${>}$~0.7~mag (solid line, 295
stars). The inset displays the spatial distribution of the sample sources with H-Ks 
${>}$ 0.5 mag within the entire area of 1.23$^\circ$~$\times$~0.64$^\circ$. \label{correlation}}

\figcaption[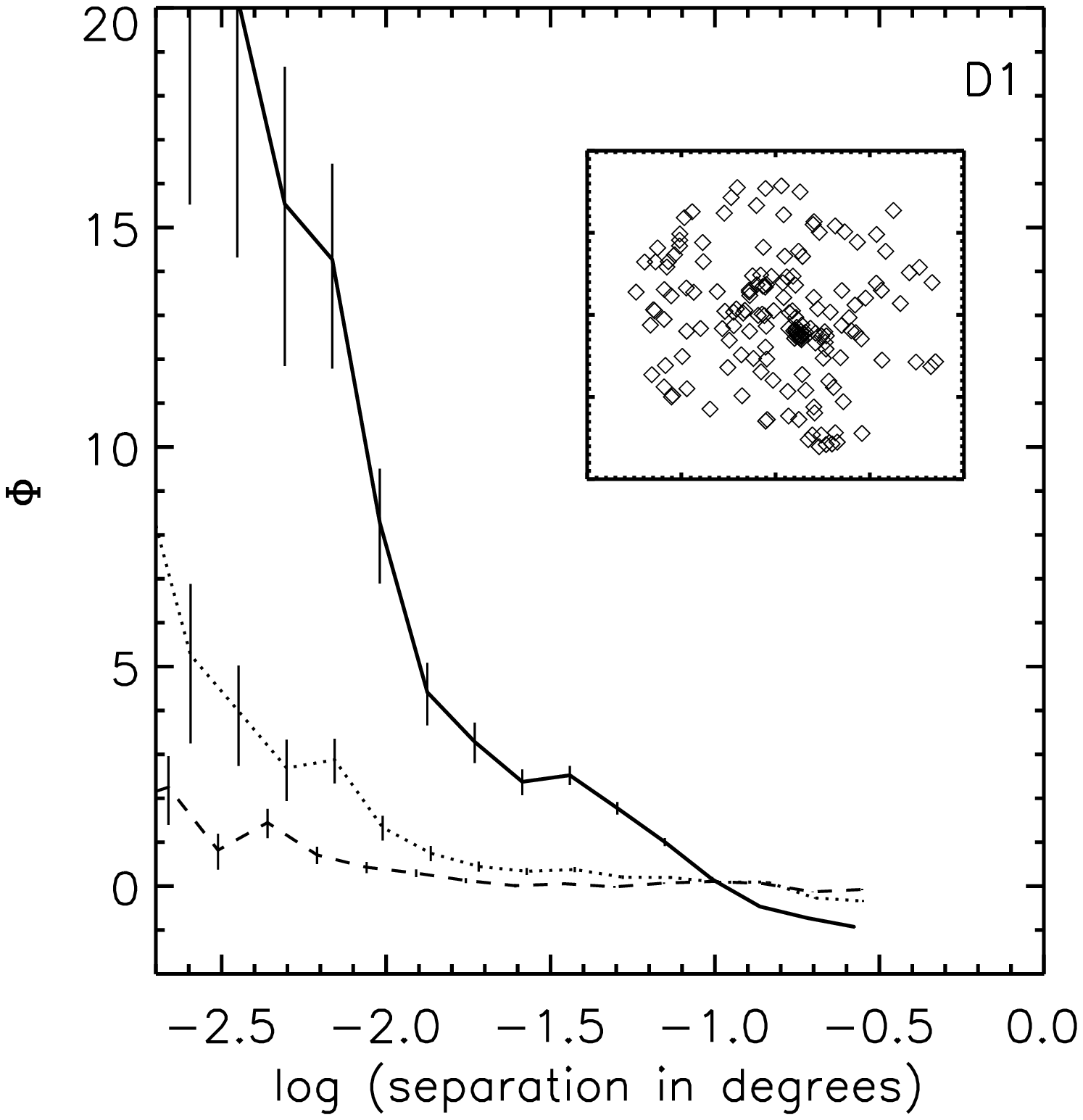]{The two-point correlation function for objects in the circular 
Sub-Region D1 of the RMC.
The three lines correspond to the data with  H-Ks~${>}$~0.2~mag (dashed line, 313
stars), H-Ks~${>}$~0.5~mag (dotted line, 166) and  H-Ks~${>}$~0.7~mag (solid line, 74
stars). The inset displays the spatial distribution of the sample sources with H-Ks 
${>}$ 0.5 mag within the area of diameter 0.32$^\circ$. \label{correlation1}} 

\figcaption[f4a.ps,f4b.ps] {The log($\Phi$)-$\theta$ correlation for Region~D (upper panel)
and Region~D1 (lower panel). A power-law correlation function is thus
linear here. The dot-dashed straight lines correspond to power-laws of the form
$\Phi \propto \theta^{-0.84}$ (upper panel) and $\Phi \propto \theta^{-0.95}$
(lower panel). The full and dotted data lines correspond to the definitions
employed in Figs.~\ref{correlation} and \ref{correlation1}. \label{corrlog}}

\figcaption[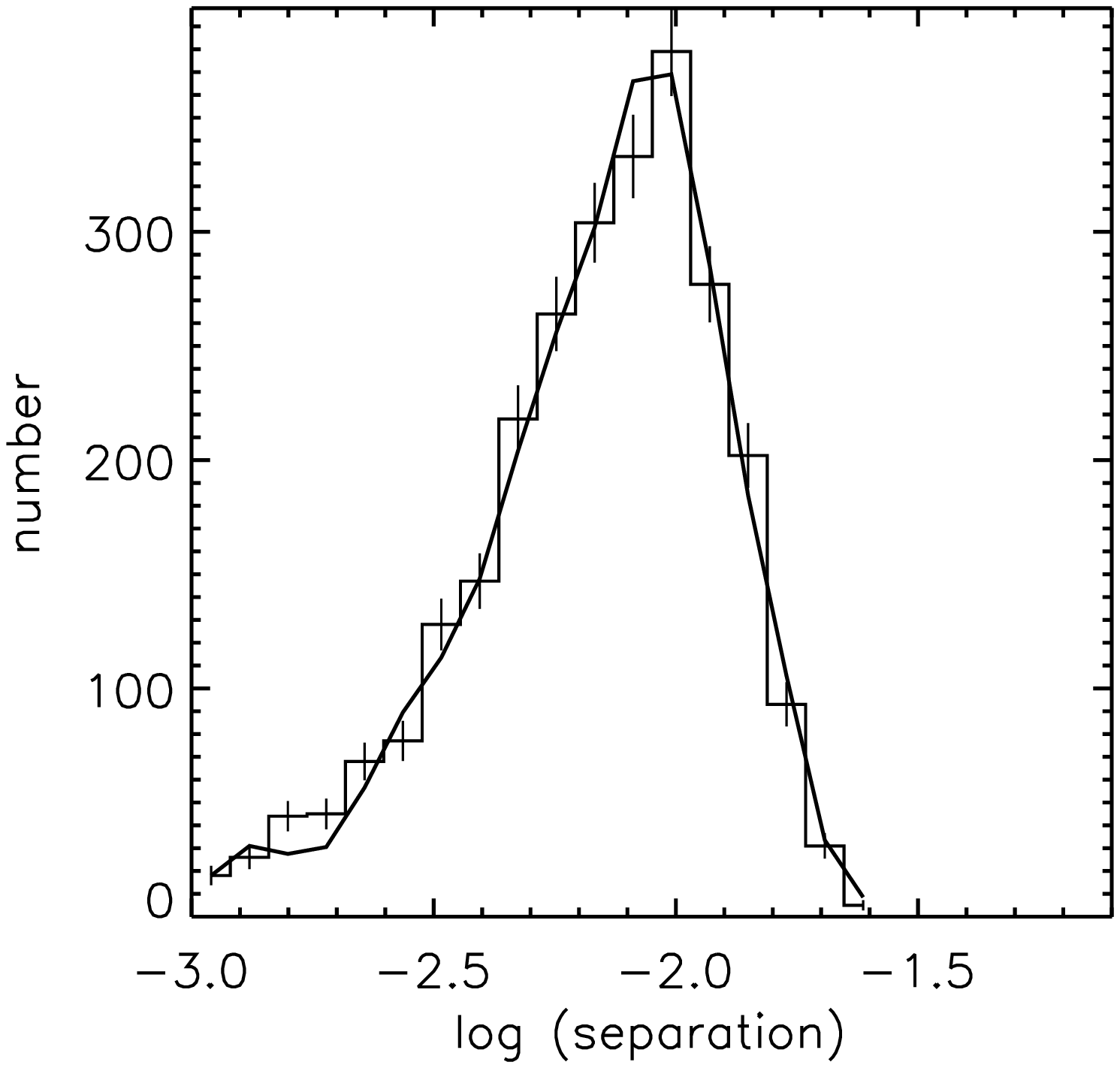,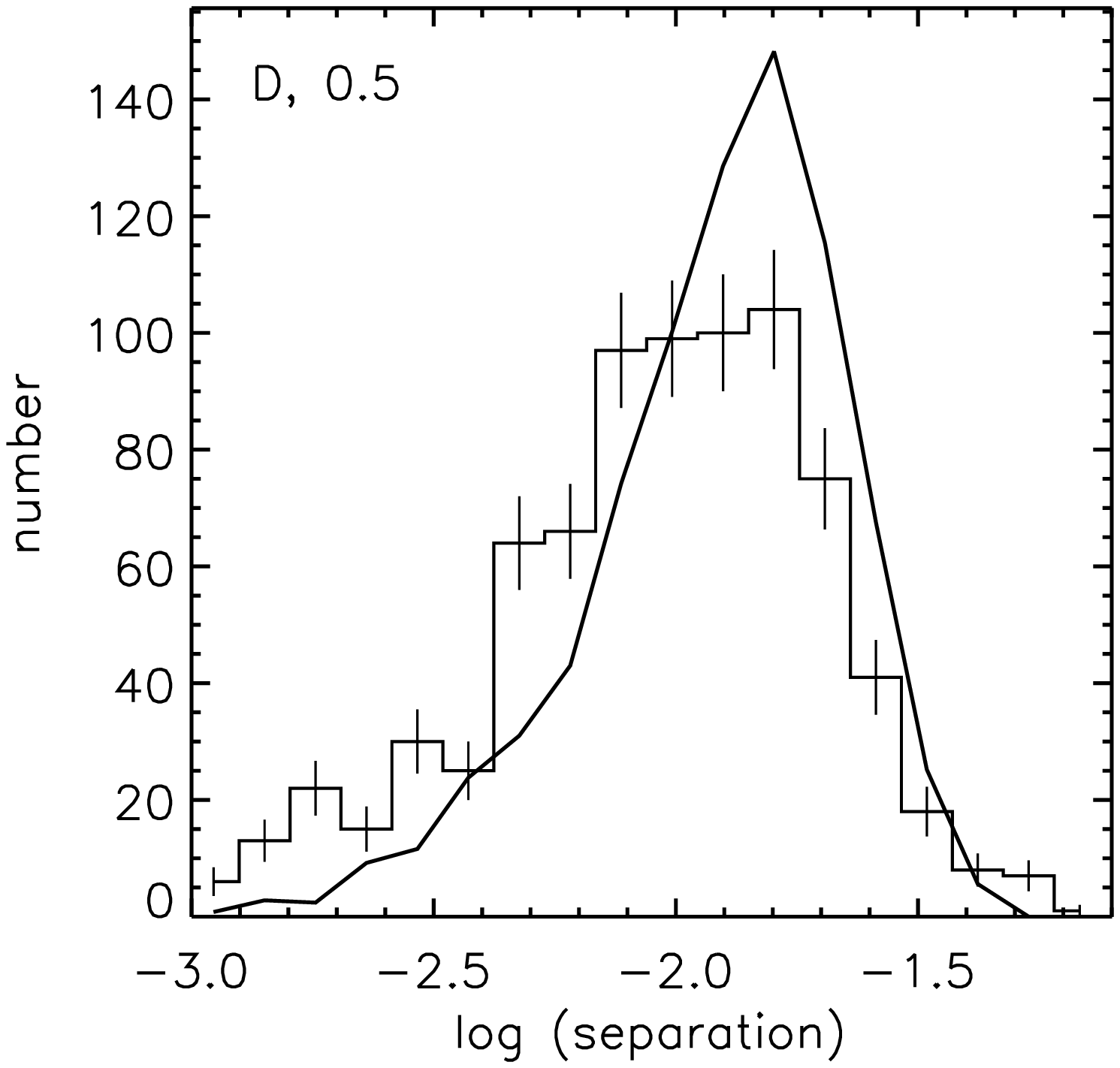,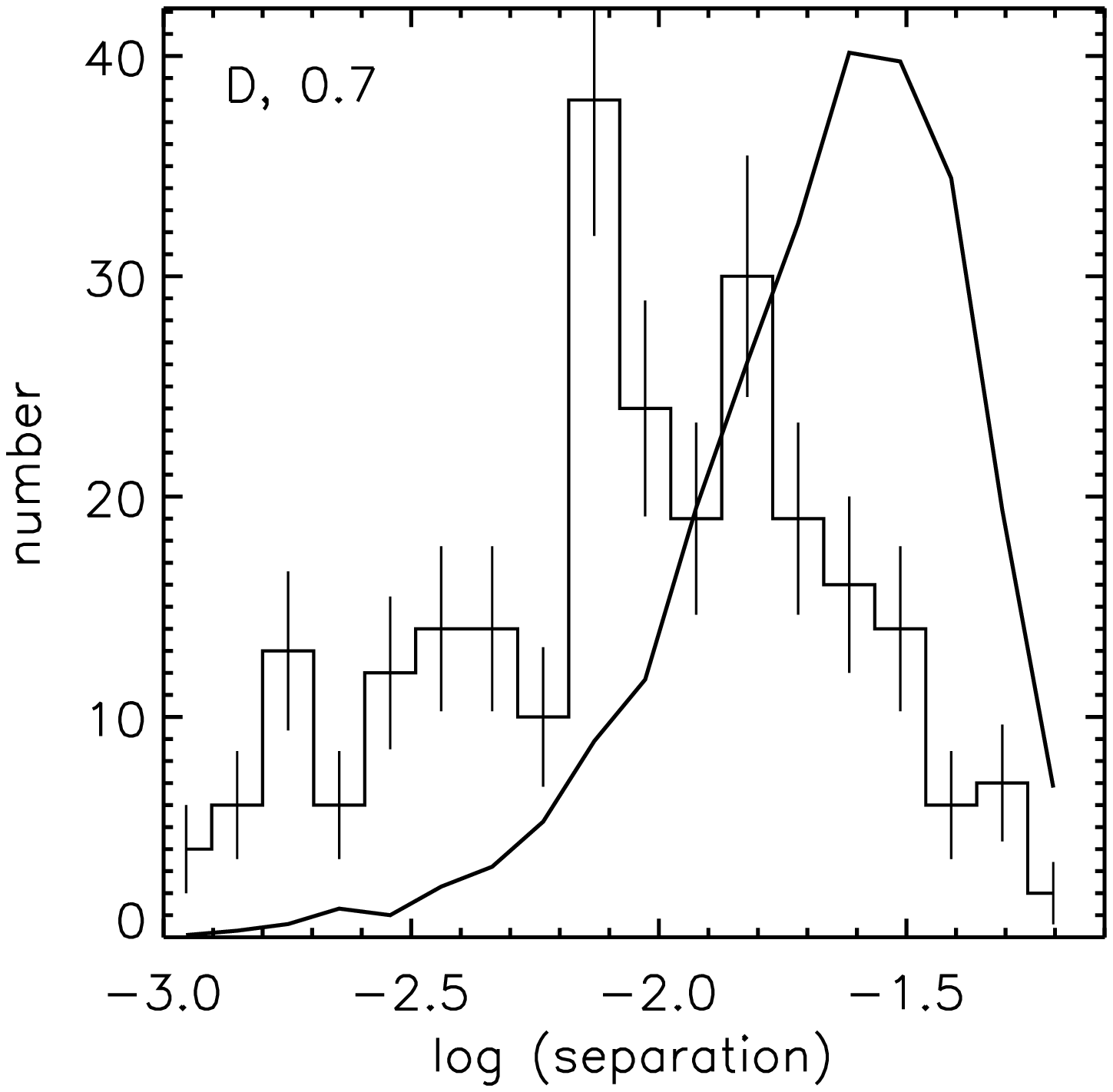]{The number distribution of
the projected separation distance between nearest neighbours for Region~D. 
As indicated, the panels correspond to the star samples with H-Ks~${>}$~0.2~mag (2660
stars), H-Ks~${>}$~0.5~mag (792) and  H-Ks~${>}$~0.7~mag (295 stars). The solid lines are
the equivalent distributions for the same numbers of randomly distributed stars within 
the same area. Error bars are calculated from square-root number statistics.
\label{neighbours}}

\figcaption[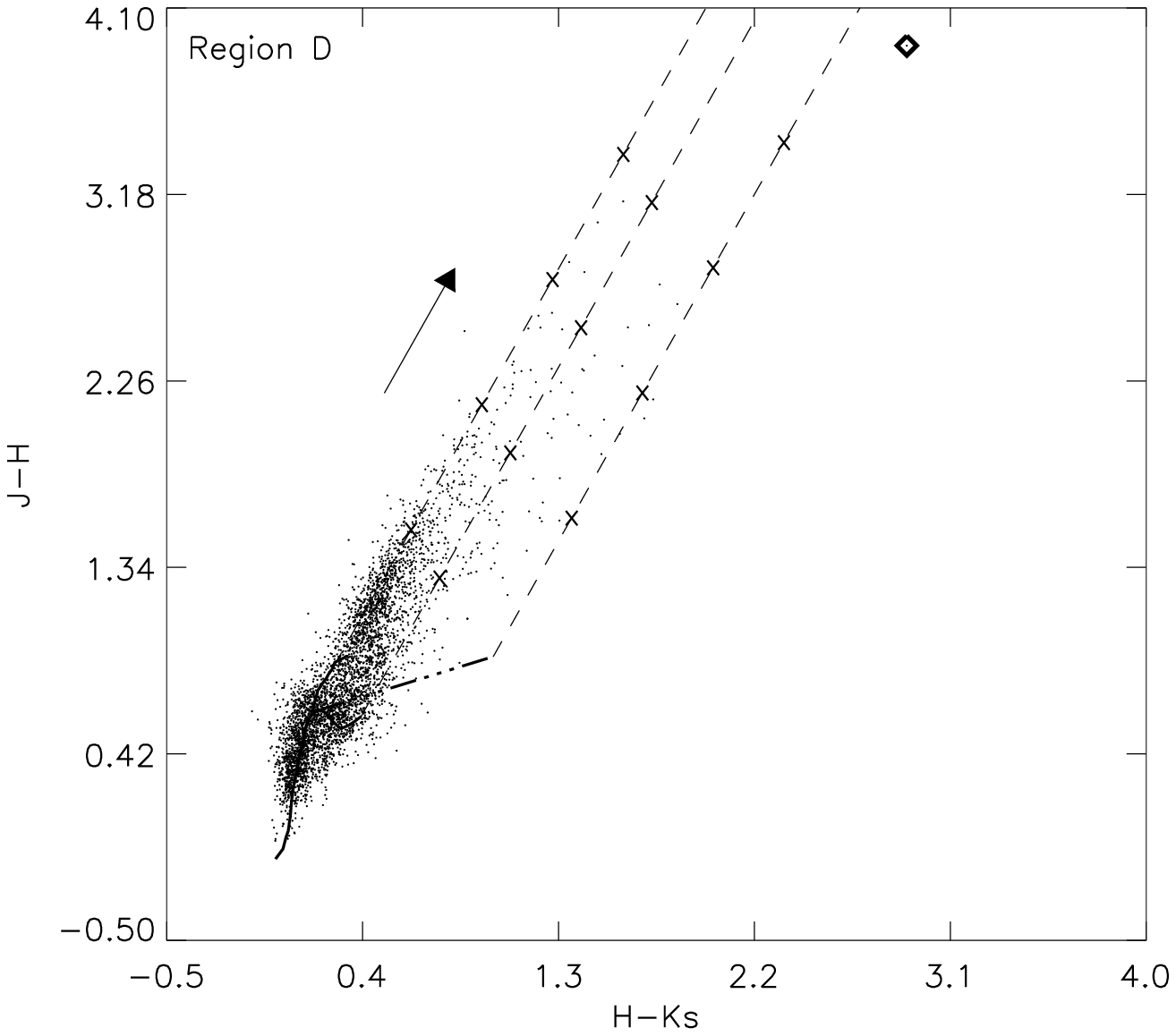,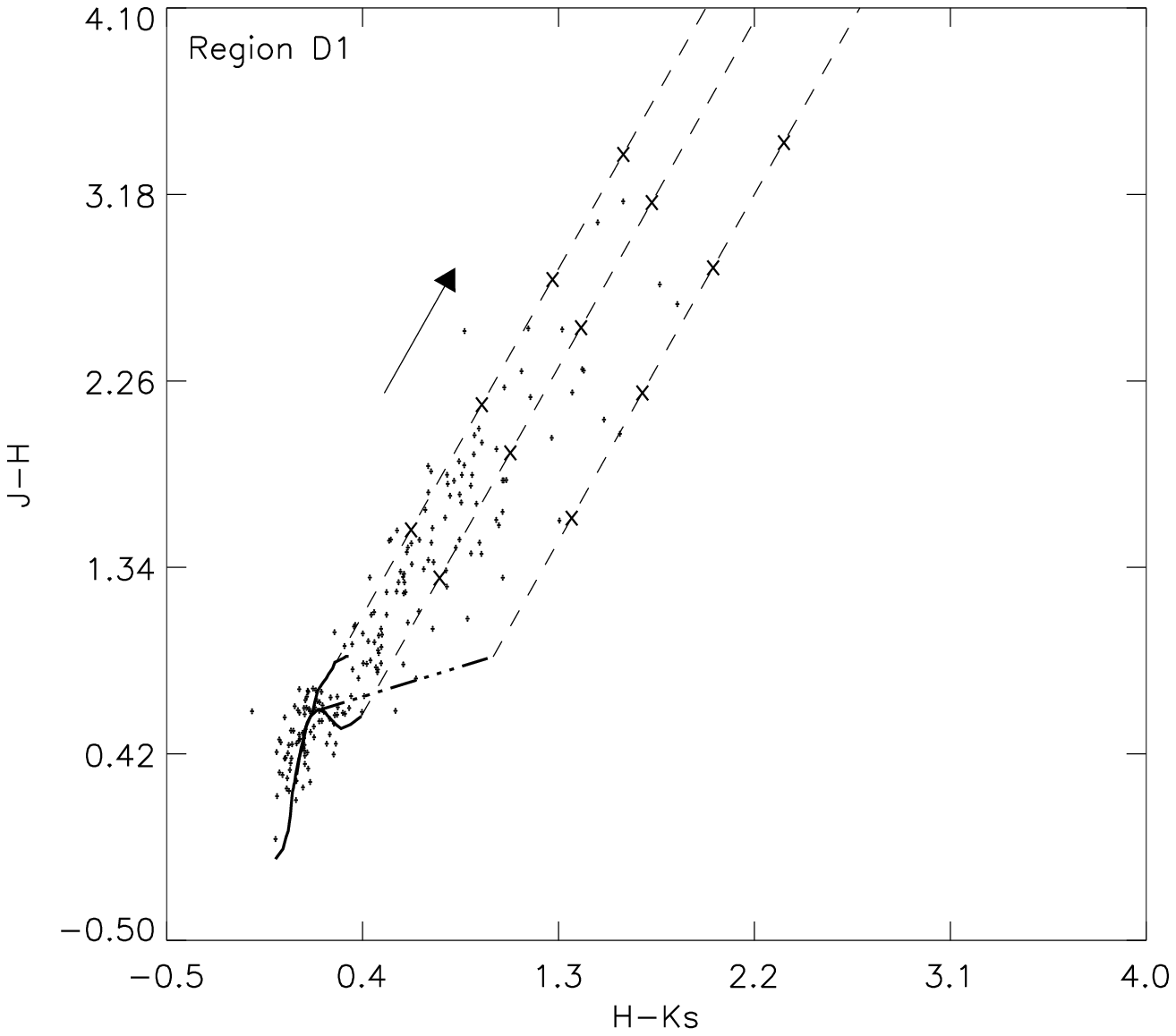]{Color-color diagrams of the Region~D cluster and its sub-cluster D1. Candidate
cluster and sub-cluster members are denoted as dots. The source with the
highest H-Ks in the Region~D diagram is indicated by a diamond. Solid lines represent the
loci of the main-sequence dwarfs and giant stars \citep{1988PASP..100.1134B}. The
arrow in the upper left of the plot shows a reddening vector of Av~=~5\,mag
\citep{1985ApJ...288..618R}. The dotted dashed line indicating the locus of
de-reddened T Tauri stars \citep{1997AJ....114..288M}. The dashed lines define the
reddening band for the normal stars and T~Tauri stars, and are drawn parallel
to the reddening vector; crosses were overplotted with an interval corresponding
to 5\,mag of visual extinction. \label{colorcolor}}

\figcaption[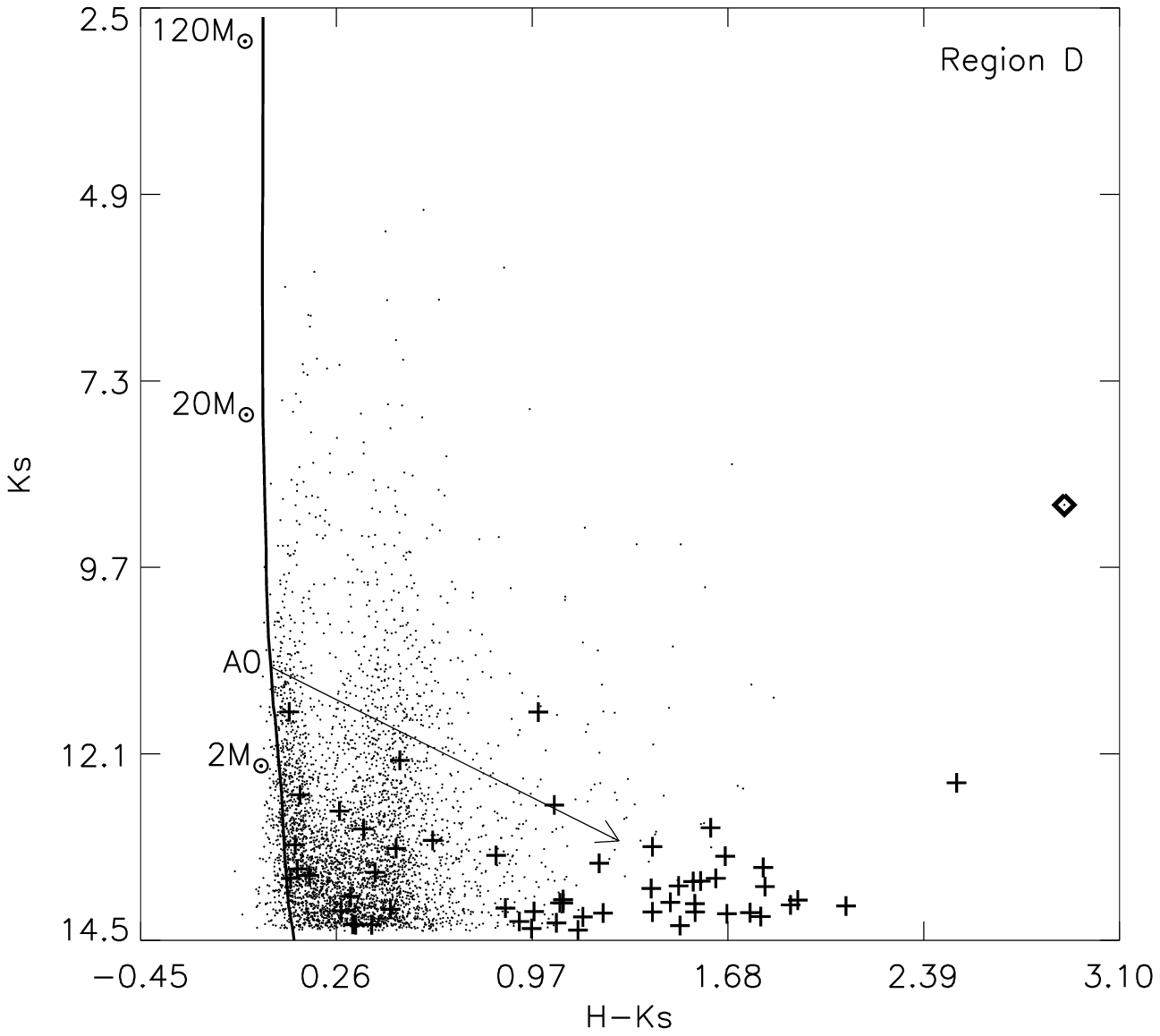,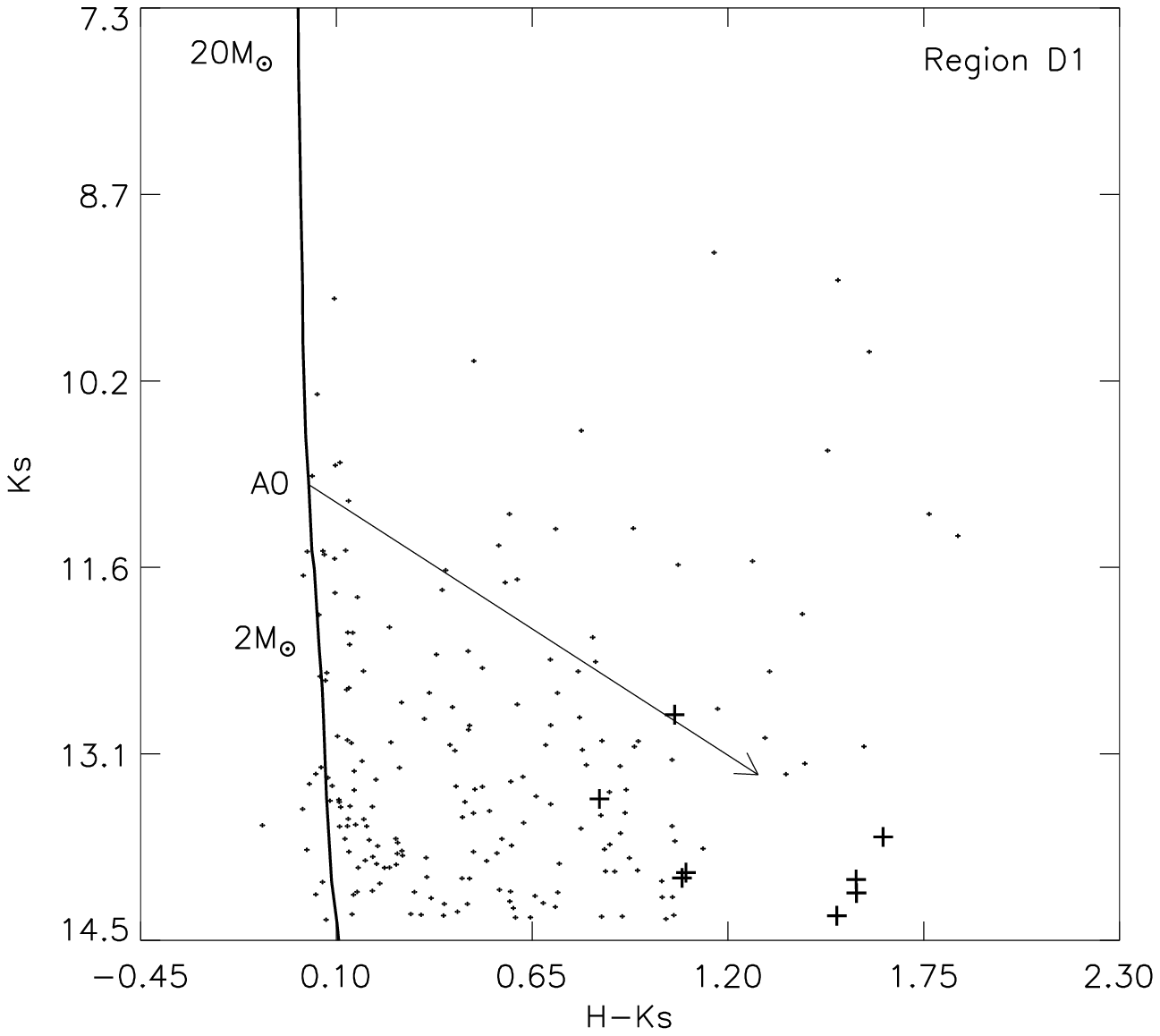]{Color-magnitude diagrams of the Region D cluster and its sub-cluster D1.
Source indicators are the same as Fig.~\ref{colorcolor}. Sources missed by
the selection criteria are plotted with plus signs. The unreddened main-sequence 
is plotted as a solid line \citep{2001A&A...366..538L}. The slanted line with an
arrow at the tip denotes a reddening of Av~=~20\,mag of an A0 type dwarf.\label{colormag}}

\figcaption[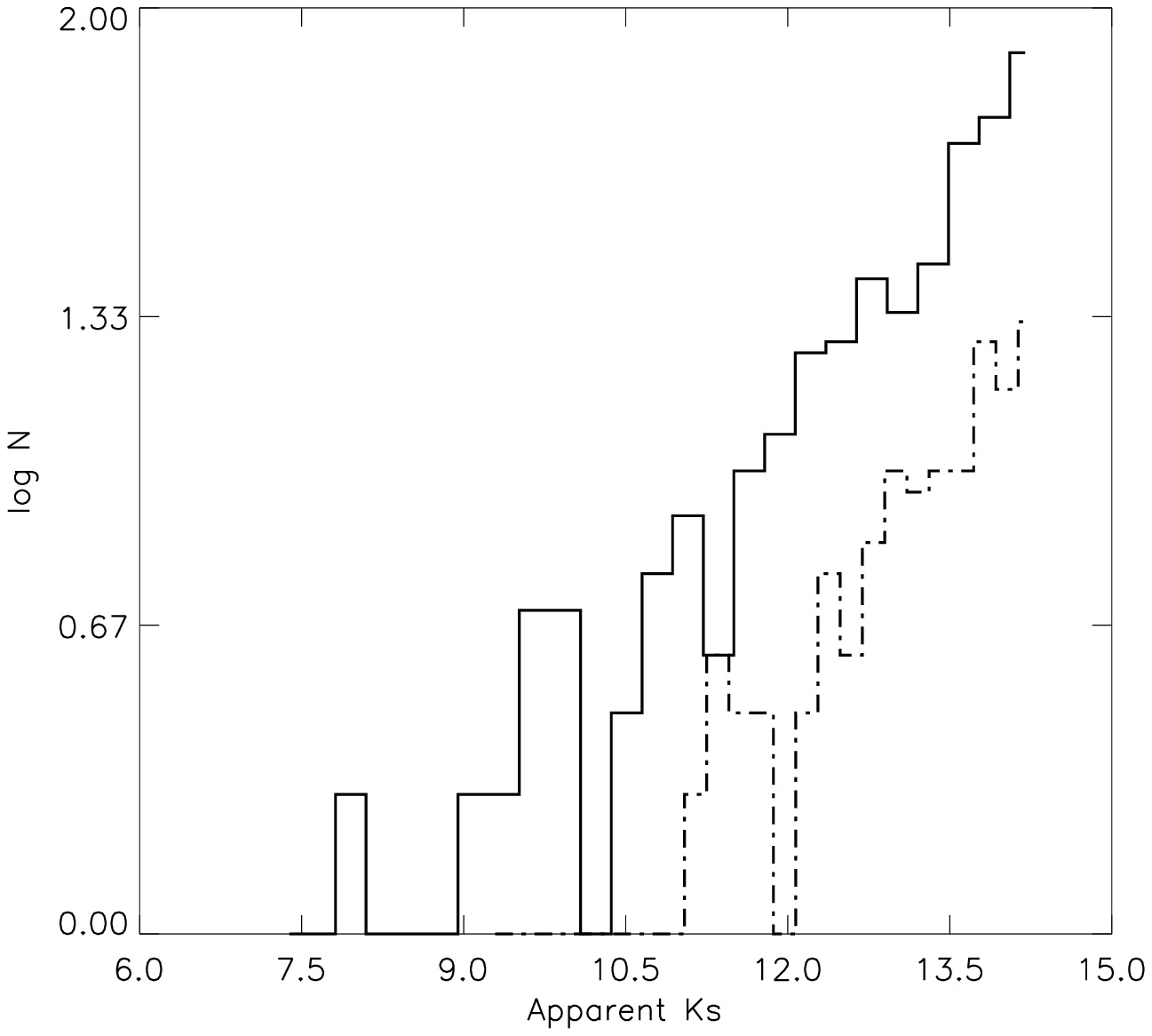]{The KLF of the Region D cluster (solid line) and its sub-cluster D1
(dotted line). It is evident that both Regions D and D1 have
a very similar slope of $\sim$~0.4 in the range of 11.0 ${<}$ Ks ${<}$ 14.5 mag,
signifying that the embedded sub-clusters have ages of only $<$~1\,Myr and 
are still engaged in their early stages of evolution. \label{lumin}}

\figcaption[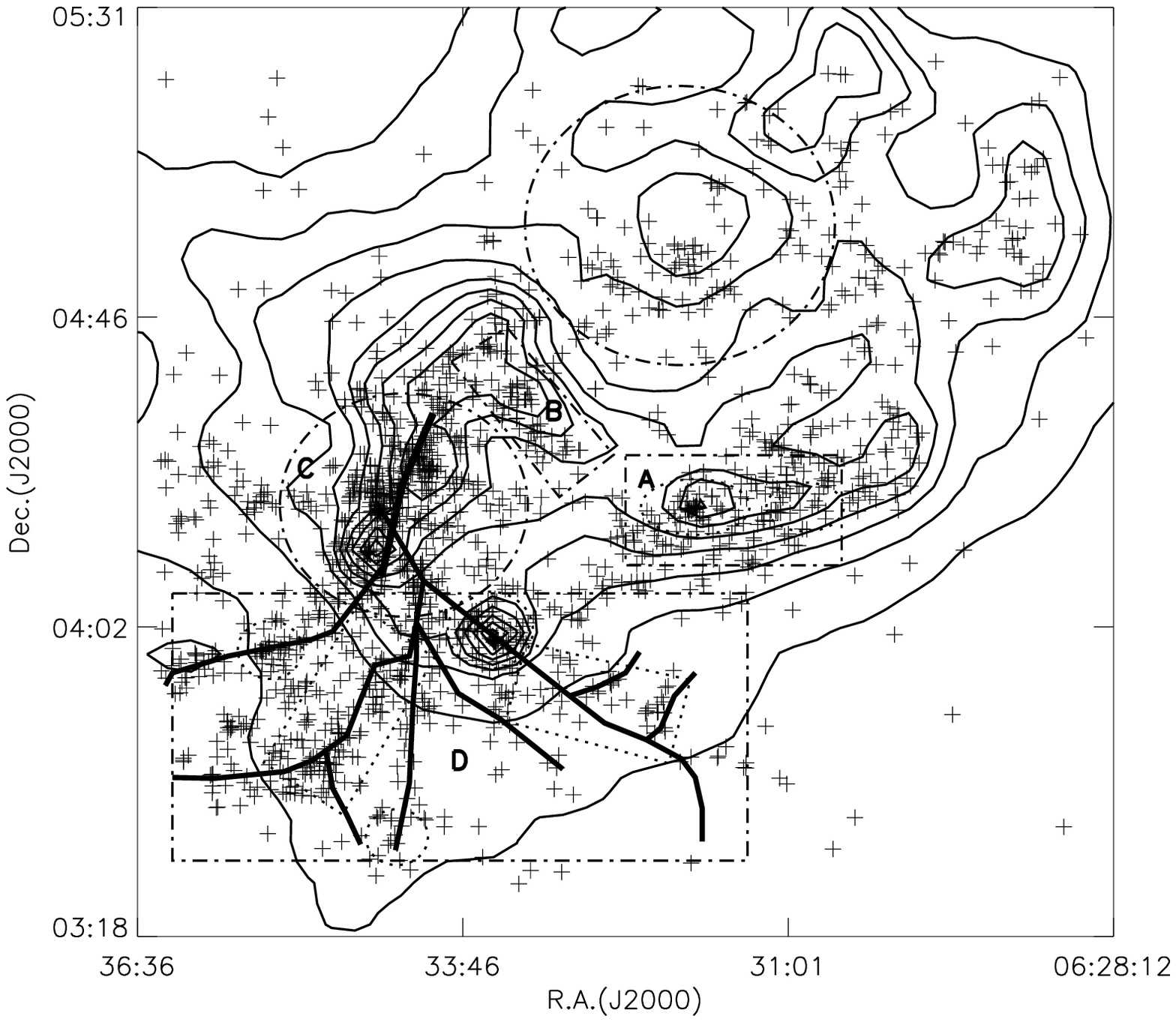,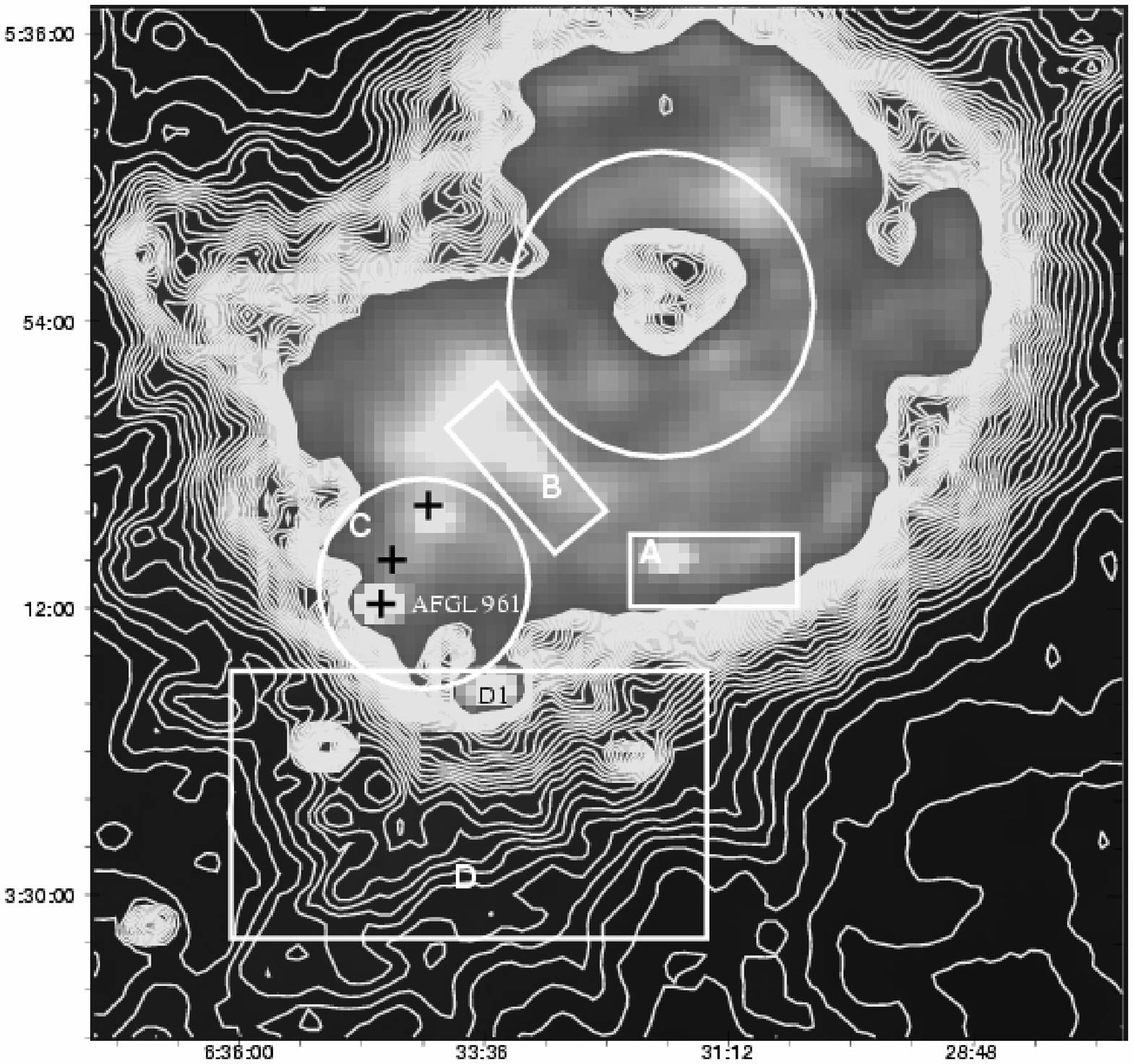]{Upper panel: The structured distribution of the embedded clusters and
loose aggregates of candidate young stars in the entire RMC. 
A tree pattern is schematically illustrated by the thick continuous lines overlaid. 
Lower panel: An intriguing distribution of the IRAS 60\,$\mu$m emission of the RMC that 
matches quite well the star formation branches. \label{structure}}

\clearpage
\epsscale{0.85}
\plotone{f1a.ps}
\plotone{f1b.ps}
Fig.1
\clearpage
\plotone{f2.eps}
Fig.2
\clearpage
\plotone{f3.eps}
Fig.3
\clearpage
\plotone{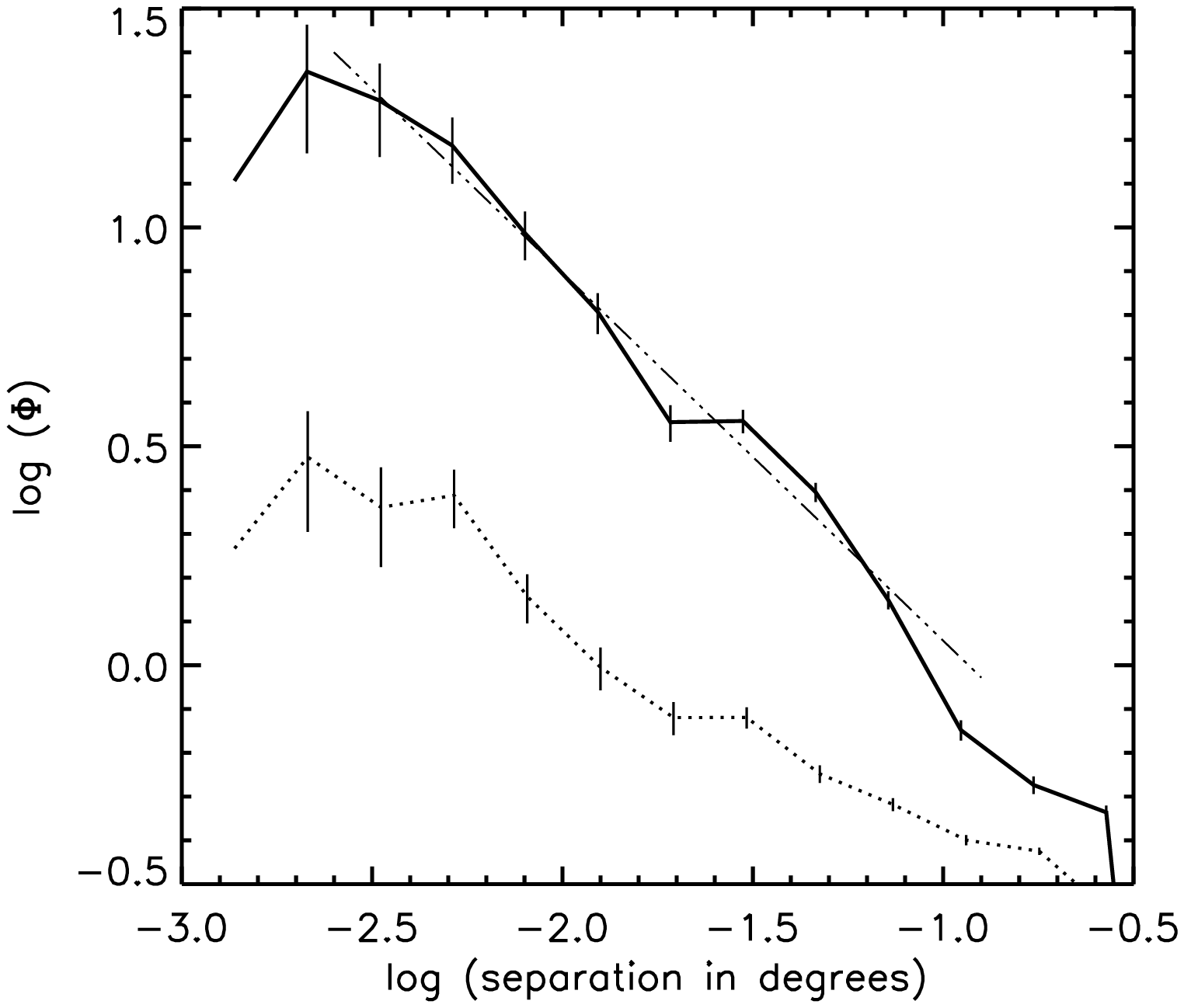}
\plotone{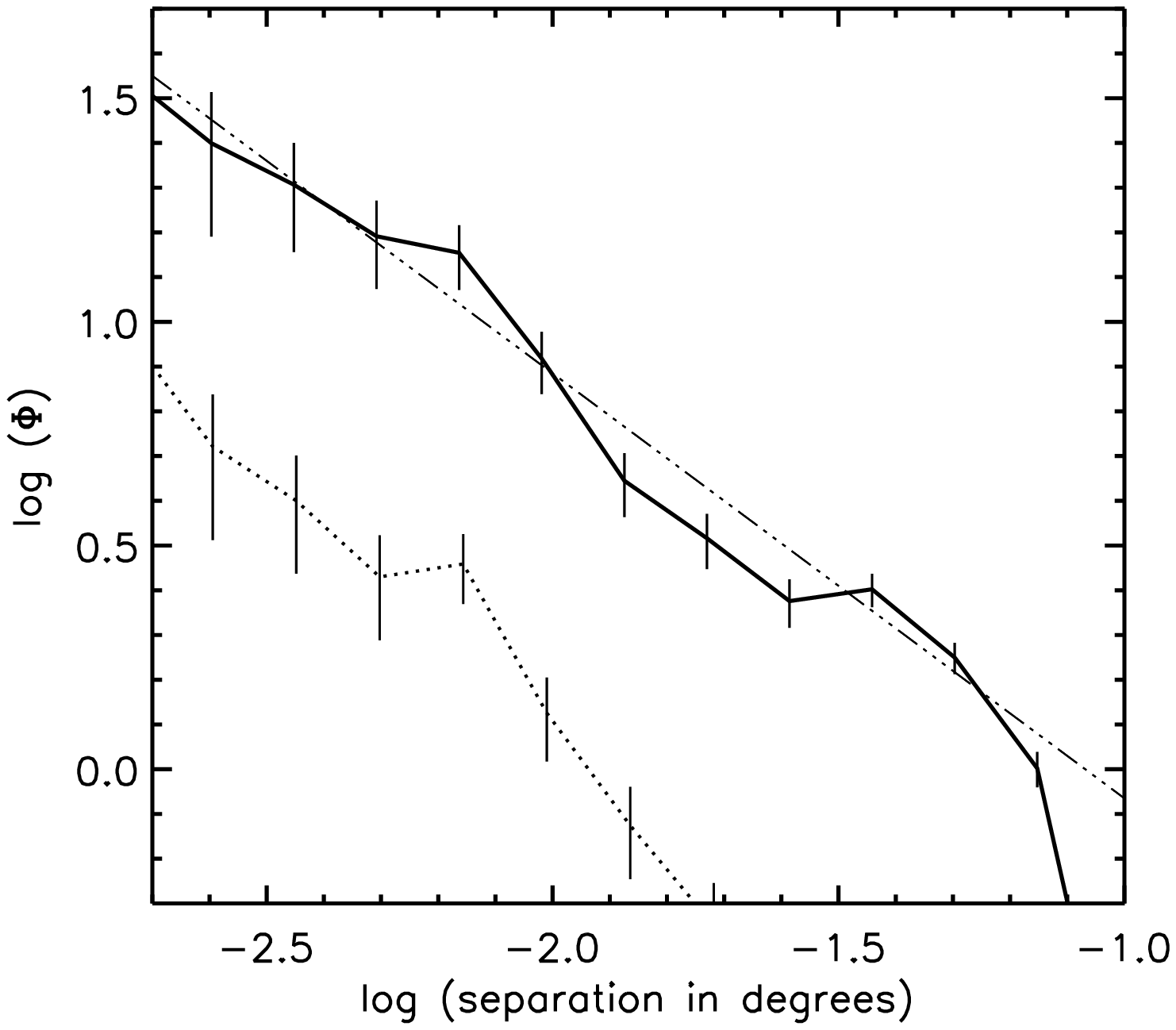}
Fig.4
\clearpage
\plotone{f5a.eps}
\plotone{f5b.eps}
\plotone{f5c.eps}
Fig.5
\clearpage
\plotone{f6a.ps}
\plotone{f6b.ps}
Fig.6
\clearpage
\plotone{f7a.ps}
\plotone{f7b.ps}
Fig.7
\clearpage
\plotone{f8.ps}
Fig.8
\clearpage
\epsscale{0.7}
\plotone{f9a.ps}
\plotone{f9b.ps}
Fig.9

\end{document}